\documentstyle[12pt,aaspp4]{article}
\slugcomment{Version %I% %D%
, To appear in Astronomical J.}
\lefthead{Kochanski, Tyson, and Fischer}
\righthead{Flickering Faint Galaxies: Few}
\newcommand{\etal}{{\it et al. }\ }
\newcommand{\eg}{{\it e.g.},\ }
\newcommand{\ie}{{\it i.e.},\ }

\begin{document}

\title{Flickering faint galaxies: few and far between}

\author{Greg P. Kochanski,
J. Anthony Tyson\altaffilmark{1}\altaffilmark{2}
\altaffiltext{1}{Visiting Astronomer, Canada 
France Hawaii Telescope.}
\altaffiltext{2}{Visiting Astronomer, National Optical 
Astronomy Observatories,
which is operated by the Association of Universities for Research in
Astronomy, Inc., under contract to the National Science Foundation.}, 
and Philippe Fischer\altaffilmark{3}\altaffilmark{4}
\altaffiltext{3}{Hubble Fellow, now at the University of Michigan}
\altaffiltext{4}{Visiting Astronomer, Canada France Hawaii telescope}}

\affil{AT\&T Bell Laboratories, 600 Mountain Ave., Murray Hill, NJ 07974}
\authoraddr{gpk@research.att.com}

\begin{abstract}

Optical variability in galaxies at high redshift is a tracer of
evolution in AGN activity, and
should provide a useful constraint on
models of galaxy evolution, AGN structure, and cosmology.
We studied optical variability in multiple deep CCD 
and photographic surveys of blank fields for 
galaxies with $B_j = 20 - 25$ mag.
Weakly variable objects are far more common than strongly variable ones.
For objects near $B_j = 22$,
$0.74\% \pm 0.2 \%$ vary by 0.026~mag RMS or more, over a decade.
This is small compared with previous claims based on photographic surveys,
and also small compared with the fraction of bright quasars
($\approx 5\%$ at $B_j = 20$~mag)
or Seyferts ($\approx 1-2\%$ for $B_j < 18$).
%%  *********************************************
%%  1 QSO/sq.deg/mag at 18 Bj mag. == 3% of gal dens
%%  at 19th QSO fraction = 7% of gal dens
%%  *********************************************
The fraction of objects that vary increases slowly with magnitude.
Detection probabilities and error rates
were checked by simulations and
statistical analysis of fluctuations of sample sky spots.
%
% It seems likely that most quasars are not substantially variable
% (\eg 0.3 mag) on 1-10 year time scales.
% However,
% small amounts of variability are likely to be present in many quasars.

\end{abstract}


\section{Introduction}

The existence of Quasars (QSOs) near redshift 5
(\cite{SCHa89}, \cite{SCHb89}, \cite{SCHN94})
% (Schneider, Schmidt and Gunn, 1989 and 1993)
and the recent discovery of a rapidly rotating 0.1~pc disk in a galaxy
(\cite{MIY95})
%(Miyoshi \etal 1995)
suggests that there are many
black holes that predate
galaxies and have accretion disks that
vary
noticeably over days to months
(\cite{TUR91}, \cite{LOER94}, \cite{EIS95}).
%(Turner 1991; Loeb \& Rasio 1994; Eisenstein \& Loeb 1995).
Eisenstein and Loeb argue that more than 0.1\% of early objects with baryonic
mass in the range $10^6$--$10^7$ M$_\odot$ have very low angular momentum
and would settle within $10^6$ yr to a $\sim$ 0.1~pc compact disk, quickly
evolving into a seed black hole.  
Given these seed black holes, one expects the seeded galaxies to
have AGNs (Active Galactic Nuclei) and to exhibit optical variability when the
black holes are accreting.

%Pre-formed compact galaxy nuclei would provide a mechanism for early AGN
%activity, given adequate accretion.

Models of the origin and evolution of activity in galaxies are constrained
by the number density and variation amplitude of such galaxies.
While variation of the optical luminosity of AGNs is known to exist
on timescales of days to years,
the AGN phenomenon has traditionally been studied via
other indicators of activity such as line, radio, and X-ray emission
%(Balick and Heckman 1982; Rees 1984).
(\cite{BAL82}, \cite{REE84}).
Little is known of optical variability in unbiased samples of local galaxies,
much less at high redshift where evolution of this phenomenon might be detected.
Searches for early AGN activity via optical variability are a good way to
investigate early compact objects,
as variability over a 1~-~10~year period requires a compact
(\ie less than 10~pc) source,
such as a massive central black hole in a galaxy.
The source must also not have much local obscuration,
as light which has been scattered through large angles
(thus randomly delayed)
will have had its short-term variability smoothed away.

Large CCDs and gigabyte disks enable an accurate search for these effects.
In this paper we present the results of high precision photometric
monitoring of 2830 galaxies in a single $16\arcmin^2$
field (2345+007) over 13 epochs spanning 1984 to 1994.
The mean magnitudes of the galaxies are
$B_j = 24.8$ mag and $R = 23.3$ mag.
We also compare our CCD data to photographic data from
a wider $1.16\arcdeg^2$ field,
whose objects have a mean magnitude of $B_j = 23.7 $ mag.

\section{Observations}

The 2345+007 data used in this study were taken during
on the CTIO 4-m, KPNO 4-m, and CFHT 3.6-m telescopes,
over the period 1984 -- 1994.
A typical epoch consists of deep
shift-and-stare exposure sequences
%(Tyson 1990)
(\cite{TYS90})
in $B_j$ and $R$ filter bands,
reaching $B_j \approx 26$ magnitude
($26.5 B_j$ % CT1092JbT sky noise from focas_sky
and $25.5 R$ mag arcsec$^{-2}$	% CT1092RbT
surface brightness $3 \sigma$ threshold).
The size of this survey field has increased with time from 
$2.5 \arcmin \times 4 \arcmin$ to $16 \arcmin \times 16 \arcmin$.
Table \ref{journal} is a journal of these observations.
Figure \ref{color} [Plate 000] shows a composite color image of the central
area of the field.
In this image, the frames have been combined with a
weighted clipped average
%(Fischer \& Kochanski 1994).
(\cite{FIS94}).

These images, and other ``blank'' field data taken during each run were used
to generate a night sky superflat.
The superflat was used to process the data,
then all frames in a given color for a given epoch
were aligned and combined by a soft-clipped average.
The final images for each epoch are essentially free of radiation events
and CCD defects.
Calibration was obtained from transfer standards within this field
%(Tyson \& Seitzer 1988)
(\cite{TYSb88})
referenced to fundamental standards for the
$B_j$ \& $R$ system
%(see Gullixson \etal 1995).
(\cite{GUL95}).
For the purposes of this study we
require only relative photometry; the brightness of each
object is compared to the average of all other objects in the field.
Relative photometric accuracies range from 0.01~mag to 0.05~mag
for bright objects
(see section \ref{photometry}, and figures \ref{photerrG} and
\ref{photerrB}).

As a check on our CCD survey,
we also include data from
a wider area photographic survey in Sextans with only two epochs.
The photographic data consists of
48000 galaxies of 19~--~23~$B_j$~mag in a
$1.16 \arcdeg \times 1.16 \arcdeg$,
two epoch survey from the 2.5m Dupont telescope at
Las Campanas Observatory
% (Tyson 1995; Postman \etal 1995).
(\cite{TYS95}, \cite{POS95}).
Six of nine plates were used from a run on March 14-23, 1985,
and three of four from a run on march 13-15, 1986.
Exposure times were two hours on hypersensitized IIIa-J plates;
the seeing was $1.2\arcsec$ FWHM on the combined images. 
The plates were scanned with $0.54\arcsec$ pixels,
then density-to-intensity conversion was done on a per-pixel basis
using CCD calibrations in five sub-fields within this large field.
Thirty-four stars were used in the (nonlinear) calibration,
with a RMS residual of 0.12~mag.
Finally, the images were median-combined to reduce plate defects.
Number counts of galaxies in this field are complete to $B_j = 25$.

\section{Data Processing}\label{dataprocessing}

The goal of this program is to determine the fraction of photometrically
varying galaxies in the field
as a function of magnitude and of the degree of variability.
Relative epoch-to-epoch photometry is
complicated by variations in seeing, CCD pixel size, exposure times,
alignments and image scale between different epochs.
In order to
overcome these difficulties, we have developed software which performs
pairwise comparisons between the overlap regions of different epochs. For
each pair of observations, the individual epochs are geometrically
transformed, convolved and scaled using a least-squares fitting algorithm,
followed by the subtraction of one from the other.
Photometry is done on the subtracted images, so that the diffuse parts
of the galaxies (which are constant) cancel, and only the variable,
unresolved nuclei appear.

For 2345+007,
eighty-eight of the difference images were used, the software selects
$\approx N log_2(N)$ of the $N (N-1) /2 $ images, somewhat arbitrarily,
to keep the processing time reasonably small.
The 2345-007 and Sextans datasets are treated independently until
the last step,
where a statistical description of galaxy variability
is derived from a maximum likelihood estimator.

Objects in 2345+007 were defined by two {\it FOCAS}
%(Jarvis \& Tyson, Valdez)
(\cite{JAR81})
detections on the same epoch
(\eg $B_j$ and R), or three detections in different epochs.
For the photographic dataset, a detection in both epochs was required.
Requiring that the object be present in both images
eliminates hundreds of nominally variable ``objects'' from
consideration that are apparent on only one image.
Oversized areas were excluded for diffraction spikes and saturated
regions of bright objects, edges, and obvious plate or CCD defects.
In the Sextans field, this resulted in elimination of all objects
brighter than $\approx 19$ mag.

\subsection{Image Subtraction}

The image subtraction program, called {\it NLSFIT}, was written in C++, and
built around a standard nonlinear least-squares fitting routine.
The strategy is to fit one image to the other, in a
model-independent manner,
without the necessity of extracting
and fitting luminosity profiles and positions,
or pre-classifying objects as stellar vs. nonstellar.
It provides precise matches between images
without using isolated reference stars,
such as deep images or crowded fields.

One advantage of our approach (subtraction followed by photometry)
is that it is very insensitive to crowding effects.
Since few objects are variable, neigboring objects will subtract
out and not disturb the photometry on the difference image.

Before subtracting the images,
we must adjust the point-spread functions of the images.
Simply convolving one image with a smoothing kernel before subtraction to
match point-spread functions (PSFs) is not appropriate;
least-squares fitting algorithms assume
that the errors in different data are uncorrelated.
That assumption is violated by convolving just one dataset with a kernel.
Because of this, a naive implementation will always over-smooth.
Consider a single object in the midst of a large area of sky.
The errors in the immediate vicinity of the object will be minimum
when the smoothing is correct, \ie the PSFs are matched.
However, the noise in the large area of sky will monotonically decrease
as the smoothing is increased.
The minimum of the overall error will be with the object oversmoothed,
because
the extra systematic mismatch introduced
in the objects by oversmoothing will be compensated by a reduction of
the random noise in the blank sky between the objects.

The cure for these problems is to fulfill the formal requirements of
the least-squares fitting algorithms.
One must convolve the
subtracted data by a kernel chosen
to de-correlate or ``whiten'' the noise,
so that errors on different pixels are independent of each other.
This whitening kernel is typically a weak sharpening kernel,
and when combined with blurring the sharper of the two images,
results
in an output image with a size intermediate between the sizes of the
two input images (see figure \ref{sumimg}).
Appendix A describes the algorithm in more detail.

Figure \ref{diffimg} [ Plate 000] shows the $B_j$ band subtraction of the
1987 and 1990 epochs in the 2345+007 field,
over approximately the same area as shown in figure \ref{color} [Plate 000].
The mildly oscillatory nature of the kernels (insets in the figure)
is apparent around some of the brighter stars.
Figure \ref{sumimg} [Plate 000] shows the sum image corresponding
to figure \ref{diffimg}.
The inserts are grey-scale representations of the two convolution kernels
that are used to prepare the sum and difference images.

\section{Photometry}\label{photometry}

After subtraction,
we carried out photometry on the difference images,
with an aperture weighted by a
Gaussian with the same size as the PSF (for AGN searches)
and three times that size (for SN searches).
The larger aperture for the SN search allows detection
of supernovae outside the host galaxy's nuclear region,
and also provides a consistency check of the small-aperture AGN search.

The sky level for each measurement was derived from an annulus
of pixels around the object of interest.
The annulus extends in radius from 2 to 4 times the aperture FWHM;
this was large enough so that sky noise is negligable
compared to noise in the object.
We excluded 50\% of the annulus to protect against contamination
from nearby objects.
This was done by taking
objects in and near the annulus,
assuming a crudely modeled ``average'' galaxy of the appropriate
magnitude sits on each site,
and then taking the half of the pixels that have the smallest
modeled contamination.
Essentially, this punches out larger disks around brighter objects,
without biassing the following mode operation against isolated
bright pixels.

% The sky pixels were then sent to a finite-width mode operation
% to derive a sky value.
% This operation simply finds the value of $p$ where the interval between the
% ${p-10}^{th}$ and ${p+10}^{th}$ percentiles is smallest,
% then takes the data value of the $p^{th}$ percentile to be the mode.
% This essentially equivalent to a conventional mode
% operation on integer data, if the quantization interval is
% chosen appropriately for the conventional mode.
% However, this finite-width mode is well-behaved for floating point data,
% and it can be easily tuned to improve the tradeoff between noise
% (noise is higher with a narrow interval),
% and contamination by objects (with a large interval, it approaches
% a median).
For sum images, a finite-width mode operation is used to derive a sky
value.
For difference images, since the histogram should be symmetric and
objects have been largely cancelled out,
the sky pixels are sent to a robust
average operation.
Pixels between the $15^{th}$
and $85^{th}$ percentiles are averaged; the tails are lightly chopped
to reduce sensitivity to any remaining image defects.

Corrections to the photometry are made for differences in the
filter and CCD color response between the different epochs.
This is done by requiring that luminosity differences between each pair of
epochs must be uncorrelated with luminosity or color.
These corrections can be 
significant, cancelling
systematic photometric errors of typically $0.1 ( B_j - R )$ magnitudes.
We also allow for excess noise from
neighboring bright objects. 

For bright objects in 2345+007, 
the RMS photometry errors
range from 0.015~mag for $B_j$ photometry with a larger (3$\times$PSF) aperture
to 0.032~mag for $R$
photometry with the small aperture (aperture FWHM equal to PSF FWHM).
These are averages over all images, and are ascribed to the input
images; errors on difference images would typically be $2^{0.5}$ times
larger.
We assume that these errors are residual gain errors after flat-fielding
the CCD, and that their magnitude decreases as $N^{-1/2}$ as more
shift-and-stare images are combined
(as suggested by a fit of residual errors to $N$).
Thus, the large-aperture photometry on the best $B_j$ image is assigned
errors of 0.01~mag.
Fainter objects are dominated by Poisson noise, which we measure from
the variability of blank sky spots.

The small apertures do give a somewhat larger scatter for the brightest
objects (\eg 0.026 {\it vs. } 0.015~mag in $B_j$), but are sized for
minimum noise for faint objects, where photon statistics from the sky
dominate the errors, rather than the PSF mismatches, pixelization,
and gain errors that are important for bright objects.
In the faint object limit, photometry with an aperture that
matches the PSF shape and size is optimal
%(Fischer \& Kochanski).
(\cite{FIS94}).
In addition, small apertures minimize contamination of photometry
by crowding effects.

We checked these accuracies by comparing them to the RMS scatter of
the calibration stars.
Our calibration stars have a median magnitude of  $B_j = 23.2$
and $R = 20.8$, and we get RMS residuals in calibration of
$0.02$ mag in $R$, 0.04~mag for the large $B_j$ calibration,
and $0.1$ mag for the small aperture $B_j$ calibration.
These residuals are accurately predicted by our noise model.

In the AGN search,
we assume that only the nucleus of the galaxy varies --
any light outside the central
area is diffuse light from large numbers of individual stars.
So, it is appropriate to use a small aperture even on extended galaxies
when searching for AGNs.

Since the Sextans data is from photographic plates,
the noise is a complicated function of intensity
and aperture size
%(\eg Dainty 1976).
(\eg \cite{DAI76}).
We used a phenomenological noise model obtained from the
data.
The model was calculated independently for each 1530x1530 pixel region
of the Sextans field.
The noise models calculated for different areas agreed well (within 20\%).
The photometric data was first processed with a crude
noise model,
so that $\chi^2$ was calculated for all objects against a model of
constant liminosity.
The $\chi^2$ values were binned, a median was taken in each bin,
the photometric noise was calculated from the median,
and the resulting values were fit with
\begin{equation}
Var(I) = exp(R(asinh(I/\sigma))),
\end{equation}
where $R(x) = (a_0 + a_1 x + a_2 x^2 + a_3 x^3) / ( 1 + b x )$
($a_i$ and $b$ are fitted parameters), and $\sigma$ is the sky noise.
After this procedure iterated to convergence (two passes),
the noise model accurately represents the photographic plate
noise characteristics.
Using the median prevents the model from being affected
by a few variable objects.
In practice, after we excluded objects with noticeable diffraction
spikes, we found that we had excluded all saturated objects,
and a somewhat simpler functional form would have sufficed.

Photometry errors for the small aperture AGN search (figure \ref{photerrG})
and large aperture SN search (figure \ref{photerrB})
are shown.
The Sextans data is deeper, and more precise, despite the fact that
it is photographic data.
While some of the 2345+007 epochs have long exposure CCD images,
many of the larger CCD images (table \ref{journal}) are short exposures.
The plotted curves for 2345+007, which are area-weighted averages
of the photometric errors, tend to emphasize the large area exposures.
Fainter than  approximately $22$ mag,
the errors are dominated by photon statistics
(for 2345+007), or photographic grain statistics (for Sextans).
Sextans errors rise dramatically for objects brighter than $19$ mag
as the emulsion saturates, however these bright objects have already
been removed because they contribute substantial diffraction spikes and
epoch-dependent ghost images from stray reflections.
Figures \ref{photerrG} and \ref{photerrB} are discussed further in the next
section.

\subsection{Detection of Variability}\label{detvar}

Given these photometric data, we must decide if a given
galaxy is variable,
and then if the variability is more likely due to an AGN or a SN.
To do this, two statistics are computed
for each location: a variability statistic, and a Supernova (SN) discriminator.
The variability statistic is $\chi^2$ of the brightness fluctuations,
while the SN discriminator is
the luminosity difference between the brightest epoch and the average of all
others,
divided by the noise of that difference.
A supernova will typically appear as a single bright epoch,
and thus will
have a large discriminator.
Conversely, an AGN, with its random fluctuations, will have a discriminator
in the neighborhood of two.

We now need to set the smallest possible threshold for
detection of RMS variability,
while retaining good confidence that the object really {\it is} variable.
To determine the statistical significance,
we follow a set of blank sky points through the
entire processing procedure, in parallel to the set of real objects.
We can thus measure real significance levels, rather than
assuming that the noise is exactly Gaussian.

The variability and SN statistics for the sky spots are binned by the
number of observations of each spot ($N$),
and the tail of the cumulative distribution function
(CDF) for the bins was fit to a empirical power-law function.
\begin{mathletters}
\begin{equation}\label{cdffit}
CDF( \chi^2 ) = 1 - { ( ( \chi^2 - p_{m} ) / { ( p_c - p_{m} ) } ) }^{p_x}
\end{equation}
\begin{equation}
p_{m} = c_{a1} (N-1)^{c_{a2}}
\end{equation}
\begin{equation}
p_{c} = c_{c1} (N-1)^{c_{c2}} + p_{m}
\end{equation}
\begin{equation}
p_x = c_{x1} + c_{x2}/N
\end{equation}
\end{mathletters}
where $p_{m}$, $p_{c}$, and $p_{x}$ are empirical functions
of the number of images that an object is within.
More specifically,
$p_{m}$ is a fit to the median of the statistics as a function of $N$,
$p_{c}$ is a fit to the value of $\chi^2$ where $CDF = 0.9$,
and $p_x$ is the exponent of the power law, and is fit
over the range $0.9 <= CDF < 1.0$.
The CDF is 1~minus the
false alarm rate,
\ie the probability that the observation would have
been produced by chance.
In fact, we find the tails of the distribution of $\chi^2$
to decline more slowly than would be expected,
typically as $\approx (1 + x^2)^{-2.5}$ for false alarm
probabilities near $10^{-3}$, rather than the expected $exp(-x^2/2)$.

As a check of equation \ref{cdffit}, we ran yet another set of
locations through the entire processing procedure,
from photometry to detection of variability.
This check set was chosen from a uniform distribution across the area
(to simulate the uniform density of faint galaxies),
rather than selected blank locations.
The different spatial distribution allows us to confirm that our
corrections to the noise model for proximity to bright objects are
indeed correct.
This check set returns the expected number of false variables,
within statistical uncertainties.

Figure \ref{histogram} shows the distribution
of $\chi^2$ statistics for the Sextans objects.
The difference between the statistics for the objects and blank sky
spots is apparent.
The 2345+007 histogram is similar, but has substantially fewer objects.
Objects were declared variable when the false alarm rate went below
$5 \times 10^{-4}$ for the 2345+007 dataset, and $3 \times 10^{-4}$
for the Sextans dataset.
These false alarm rates correspond to
one false claim of variability out of six variables in 2345+007
and 6 of 80 for Sextans.
We used a lower threshold for the Sextans dataset simply because we had
more data available and could afford to be more selective.
The different thresholds were consistently applied during the measurement
of the classification probabilities,
so that the final statistics (after the maximum likelihood estimator)
would be independent of the exact value of the threshold.
Objects were then classified as SN,
AGN, or uncertain, depending on the value of the SN discriminator.

Figures \ref{photerrG} and \ref{photerrB} displays photometry errors
and variability-detection thresholds versus magnitude.
Figure \ref{photerrG} displays errors on the small aperture
AGN search,
and figure \ref{photerrB} the large aperture (SN search).
The thresholds for detection of variability are $3-6$ times higher
than the $1-\sigma$ photometry errors,
and are shown by the cloud of dots, one point per object.
In general, each object has a unique detection threshold because
of three factors.
First, the noise is a function of brightness.
Second, in the 2345+007 dataset, the noise will differ from region to
region,
depending on which images (which epochs) cover a certain region.
Finally, the few points that are near other objects have increased
noise because changes in the PSF from epoch to epoch change the
blending of the images.

Almost all the objects that were seen to be variable
show a fairly small variability,
the vast majority having $\delta M < 0.3$ mag.
In fact, most of the detections are not far above the respective
threshold, an observation that implies that there is no
well-defined, distinct set of variable objects.
Instead, we are sampling the tail of a 
broad distribution of variability, which may encompass
all galaxies. 

\subsection{Simulations}

Finally, we measured the various variability-detection probabilities by adding
simulated AGNs (Gaussian intensity fluctuations with a $f^{-1.5}$
power spectrum) and simulated SNe
(data sampled from an average type-Ia light curve
from
%Dogget and Branch (1985)
\cite{DOG85}
)
to the actual photometric results for blank sky spots.
We then counted how many of the varying objects were recovered.
We were also able to quantify the probabilities of misclassification
(\eg a SN classified as an AGN).
Figure \ref{classprob} shows the two classification
and four misclassification
probabilities for the 2345+007 dataset.
As these curves are measured by adding variability to blank sky spots,
they are the appropriate probabilities for faint objects where
the noise is dominated by sky statistics.
For brighter objects, we scale these curves by the ratio of the
object's noise level (\eg including CCD gain variations)
to the sky spot noise level.

By multiplying the variability-detection probability for AGNs by the number of
galaxies per unit magnitude,
we find that we have the most information
about galaxies with 0.1~mag variability at $B_j = 23.5$
or $R = 22.0$ magnitude.
This dataset thus provides an excellent test of Hawkins' claims
of large variability fraction at 23 $B_j$ mag (see discussion below).
Galaxies with smaller variability must be correspondingly brighter
in order to have a sufficiently large signal to noise ratio to be
clearly variable;
galaxies with 0.03~mag variability would show up predominantly
near $B_j \approx 21$, depending somewhat on the color
and aperture size.
Our variability-detection limit, for a 33\% chance of detection,
when averaged
over the $2048^2$ pixel CCD field, is $B_j = 24.5$ mag,
for an object with RMS($\delta$L)/L$ = 1$.
At this magnitude, we still have a 
a near-unity probability of seeing such variation if
the AGN or SN occurs in the central, deepest area, where all
the fields overlap.

\section{Statistical Analysis}

The goal of the analysis procedure was to determine parameters and error
bars that answer basic questions about the statistical distribution
of variability in the galaxy population.
The basic questions are: ``How many galaxies vary by X mag?'',
``does the incidence of variability change with apparent
magnitude?'', ``is the $z \approx 0.3$ supernova rate similar to
the local rate?''.
To accomplish this,
the objects and the detection and classification
probabilities were then fed into a maximum likelihood estimation
routine.

This routine takes as input the measured classification probabilities
as a function of the variability, and a model describing
how much a given galaxy varies as a function of its apparent luminosity.
We calculate first the probability (in an ensemble of universes)
that we would detect each particular galaxy to be variable.
Next, we calculate
the overall probability that (in the ensemble) we would
have detected as variable those specific galaxies that we {\it actually} 
detected to be variable.
This overall probability is the likelihood of the model -- it answers
the question ``How likely is it that {\it this} model would have
reproduced the results we actually obtained?''

The model is parameterized, and the program varies the parameters
to obtain the model with the maximum likelihood.
It then obtains error bars by inspecting how fast the
likelihood drops off as different parameters are varied
by a simulated annealing algorithm.            
Information comes both from the galaxies that are seen to be variable
and those that were not; both help to constrain the model.

We used either a seven or eight parameter minimization.
One for dependence of AGN activity on luminosity.
Two parameters to define the distribution of AGN variability
(overall amount of variability, and shape of the probability distribution).
Two parameters to define SN activity
(SN rate and mean magnitude relative to the galaxies).
Sometimes (as a test)
one parameter was added that gave the AGN variability on the Sextans field,
independently from the 2345+007 field.
Finally, there are two parameters to match the 2345+007 and Sextans
datasets (see appendix B).

\subsection{AGN}

The classification probabilities (\eg the probability
that an AGN will be detected as variable and classified
as a AGN)
contain all the information on the performance of the analysis software.
We then specify a model probability distribution for AGN variability,
and how that variability might depend on apparent luminosity.
The model contains no detailed physics,
but is simply a phenomenological description of the probability
that a galaxy will change its apparent luminosity by $\delta L$.
Then:
\begin{equation}\label{agnp_eq}
P( \delta L/L ) = C (L/L_0)^\eta \max(\delta L/L, \delta_{cut}/L)^\zeta,
\end{equation}
where $\delta$ is the RMS variability of the luminosity,
$L$ is the luminosity of the object without the nucleus,
$\eta$ describes the trend of variability with luminosity,
$\zeta$ and $C$ describe the relative scarcity of strongly variable
galaxies, 
$L_0$ is an arbitrary luminosity scale,    
and $\delta_{cut}$ is a cutoff, chosen so that
${\int_0^\infty P(r) dr} = 1$.

This parameterizes the amount of variability of galactic nuclei as
a power law probability distribution,
chosen because there is no evidence that normal galaxies
and AGNs are physically distinct populations.
For instance, the mass function of Seyfert 1 nuclei
has been measured to be a broad power law distribution that does not require
any natural separation between Seyferts and normal galaxies
%(Padovani \etal).
(\cite{PAD90}).
There is no published evidence that X-ray or radio luminosities have a
truly bimodal distribution, rather than merely classification
by an arbitrary detection threshold.
Even if AGNs and quiet galaxies were distinct populations,
the common model of AGNs as dusty doughnuts
surrounding a black hole accretion disk
%(reviewed in Antonucci, 1993; \eg Coleman and Dopita, 1992)
(\cite{ANT93}, \cite{COL92})
would give a broad distribution of apparent properties
for AGNs, depending on orientation.

It is important to note that we compare the variability of the AGN
to the luminosity of the rest of the galaxy,
rather than the total luminosity.
The total luminosity can be contaminated by an arbitrarily large
amount of nuclear light (for instance, in a typical quasar, the
nuclear light dominates the diffuse starlight from the host galaxy).
The luminosity of the host galaxy is important,
because it can be directly related to the distance and the mass of
the galaxy.
Consider a total luminosity $L_t$ in the photometric aperture
which is the sum of a nuclear component and
a normal galaxy component, $L$, due to diffuse starlight: $L_t = L_n + L$.
We assume, somewhat arbitrarily, that the average value of $L_n$ is equal
to its RMS variation (for the 2345+007 dataset),
for the purpose of calculating the luminosity of the underlying galaxy.
$L_n$ is assumed to vary with a $f^{-1.5}$ power spectrum
%(in Gopal-Krishna, \etal 1995, Mangalam \& Wiita 1993)
(\cite{GOP95}, \cite{MAN93})
and thus the normalization is somewhat different for the Sextans dataset,
with its two closely spaced epochs.
The same power spectrum leads to a smaller variance for Sextans,
as it spans approximately one year, rather than ten.

We conducted six separate maximum likelihood estimations of the astrophysically
interesting parameters.
We used objects detected (as variable) with the small aperture search only,
objects detected with the large aperture search only,
and the full search.
The full search classifies objects as supernovae or AGNs, depending
on whether they have more significant variability in the small
or large aperture
(SNe are likely to be off-center, while AGNs are not).
Maximum likelihood estimates of parameters were then made both with
the Sextans AGN density equal to the 2345+007 density,
and with the Sextans density floating free.

The raw parameters are strongly correlated.
We found it simplest to express the results in terms of the
fraction of galaxies at a certain magnitude
that vary by more than a given cutoff.
We then search for the characteristic magnitude and cutoff
that result in the strongest statement (smallest error bars)
when all six sets of estimates (with error bars) are lumped together.

For $B_j \approx 22$ objects,
$0.74\% \pm 0.2\%$ are AGNs that vary by 0.026~mag or more,
RMS, over the period of observation.
We find that the probability of a galaxy having a variability
$\delta M$ (near 0.026~mag) varies as ${(\delta M)}^{-3.3\pm 0.8}$
(\ie $\zeta = -3.3\pm 0.8$).

We find that the threshold for observing a constant fraction of
variable AGNs changes by just a factor of $1.2 \pm 0.1$ per magnitude,
or equivalently, that the fraction scales as apparent luminosity to the
$\eta = -0.68 \pm 0.3$ power.
Given that the comoving QSO density is maximum at a redshift of 2,
one might expect the optical variability in 22-25~mag AGNs would
increase somewhat with magnitude.
The faintest of these galaxies would
still be at redshift $\approx 1$,
so that by going from 22nd to 25th 
magnitude we would be primarily moving out the tail of the AGN
luminosity function.

\subsection{Supernova Probabilities}

In the maximum likelihood estimate,
we take supernova rates to be equal for all galaxies.
Ideally, we would make the rate proportional to the absolute luminosity,
but without redshift information,
we cannot discriminate distant luminous galaxies from dwarf galaxies,
nor easily compare rates with local measurements.
There are thus two parameters: the rate of type-Ia supernovae, and the
mean difference between the SN peak magnitude and the host
galaxy magnitude.  We include SN-Ib and SN-II in our calculations,
scaled by the ratios obtained from \cite{VAN94}.
%Van Den Bergh (1994).
For most of the galaxies, though,
only the brightest SNe (\ie type-Ia) can be detected.

We find that the number of SNe we have observed is small
compared to the number of AGNs.
The rate is consistent with local SN rates in the vicinity of
1 per century per galaxy (\cite{TAM94}, \cite{EVA89}).
However, with current software,
we cannot accurately measure the supernova rate in these
galaxies, as the answer is too sensitive to systematic errors
in the detection probability measurements.
The small probability of misclassifying an AGN as a SN
(Figure \ref{classprob}, compare at $B_j \approx 23$)
yields a number of misidentifications comparable
to the number of detections of real supernovae.
Our detection probabilities are relatively low for SNe because
our sampling interval is typically one year; supernovae will often
flare and fade between our observations.
Thus small errors in the classification probabilities lead to large
errors in the supernova rate.
To some extent, this is an intrinsic problem.
There is no reason why an AGN cannot simulate a supernova light curve,
especially one that is sparsely sampled and at relatively low signal to noise.

\subsection{Variability Cross Checks}

We have four independent tests of our variability detections.
First, for the 2345+007 data, we can search for variability separately
in the $B_j$ and $R$ bands, and compare results.
If the galaxies are really variable,
it is likely that there is substantial
correlation between the two bands,
so we would expect the two sets of variables to be strongly overlapping.
Figure \ref{colormag} shows the 2345+007 variable objects.
It can be seen that there is substantial commonality,
suggesting that most of these detections are indeed real.
For instance, if we consider objects where the probabilities
of detection of variability are fairly large,
such as those brighter than $23^{rd}$ mag in $B_j$,
we see that 0.6\% are detected in $R$, and 0.8\% in $B_j$.
If the detections were independent,
we would expect only 0.005\% common detections, whereas we see 0.3\%.
Clearly, objects detected as variable in one color are much more likely
than average to be detected as variable in the other color.
The simplest explanation for this is that these objects are truly variable.

As a second test we compare overall variability rates in Sextans and 2345+007.
The data are taken with dramatically different techniques,
and suffer from different defects and systematic errors.
Nevertheless, we get a consistent fraction of variable objects.
The fraction of variable objects
differs by only a factor of 1.4 (with Sextans higher)
easily within combined errors.
This discrepancy can be attributed to a deviation of the AGN fluctuation
spectrum from the assumed $f^{-1.5}$ exponent,
because the length of the two datasets is so different.
Agreement of two such disparate datasets
is supporting evidence that our analysis is operating properly.

Third, we have run a end-to-end test on the 2345+007 dataset,
introducing artificial AGNs into the input images by multiplying
corresponding areas of each epoch by a random number
with unit mean, and a standard deviation of 0.1~mag.
Since the multiplication is conducted after the sky level is
subtracted, it has negligible effect on the sky, but it makes all
objects in the chosen area variable by 0.1~mag.
We then check that the number of objects that we classify as variable agrees
with our classification probabilities derived from simulations beginning
after photometry (figures \ref{photerrG} and \ref{photerrB}).

In the end-to-end test,
there were 14 objects in the area that are bright enough to
have an expected variability-detection probability
(at a 0.1~mag variability) larger
than 1\%.
The expected number of detections is 3.8, calculated from the area-averaged
detection probabilities; we detected 7 objects as variable.
Since our test area was in the central region,
where all the images overlap,
the detection probabilities are certainly expected to be better than
the areal average, which includes large areas covered
by only three or four images.
This test is thus consistent with expectations.
The seven detected objects were among those with the top nine
precalculated detection probabilities,
thus supporting the validity of our detection probability calculation.
Additionally, all seven detections were correctly classified as AGNs.

Finally, we have also scrambled the epochs in the Sextans data,
combining frames from both epochs into two independent images
that have no time ordering.
This scrambling will convert a truly variable object,
that shows large differences between the two epochs into
one with nearly identical luminosities on both scrambled images.
On the other hand, the character of the noise on the scrambled
images will not change;
specifically truly constant objects will show the same RMS difference
between images as they do between epochs.
These temporally mixed images were processed identically to the
real Sextans dataset.
They yield variability detection rates one fourth as large as
the Sextans data combined into their proper epochs.
All these detections are spurious, and they provide an estimate
of the number of errors in the real dataset.

None of these tests are conclusive, yet overall, it seems likely
that our result is a fairly accurate measurement of the true
number of variable objects.

There are also a number of partial cross-checks performed
on various parts of the process.
For instance, the calculation of the statistical significance of
detections is checked as described in section \ref{detvar}
with a set of random sky locations.
The photometry routines used here were
checked against DAOPHOT,
by comparing magnitudes and error bars of the
two (gravitationally lensed) images of QSO 2345+007,
for each epoch; results were consistent.
As mentioned in section \ref{photometry},
we compared errors derived from our calibration stars to the
measured epoch-to-epoch scatter of bright objects, and
the overall noise model.

Images of variable objects were inspected epoch by epoch to
look for imaging problems.
In the central region of the 2345+007 image,
we checked the list of object postions
that we monitor for variability against
against {\it FOCAS} detections on the full combined image
(Figure \ref{color} [Plate 000]).
We have also added simulated SNe to the input images
in earlier versions of the software,
and checked that the number of retrieved SNe is consistent with
the variability-detection probabilities shown in figure \ref{classprob}.

\section{Discussion}

The variable (and possibly variable) objects in 2345+007 are shown in figures
\ref{closeupJ} [Plate 000] and \ref{closeupR} [Plate 000].
Objects that were classified as variable on any of the nine searches
(small aperture, large aperture, or combined) x (R, B, or combined) are shown.
The figures show the actual CCD conditions; bad pixels were
marked by hand and excluded from the analysis procedure.
Variable objects in Sextans are shown in figure \ref{Scloseup} [Plate 000].
Here, we can see the advantage of separately digitizing every photographic
plate and median combining plates.
No plate defects are visible amongst the variable objects.

Table \ref{tabvar} summarizes the properties of individual objects
in 2345+007 that are clearly variable.
Most of the objects that we see
are blue (figure \ref{colormag}),
and unresolved (figure \ref{sizemag}),
and they
could be interpreted as faint AGN, distant QSOs,
or perhaps variable stars.

We can compare these results with those of Hawkins
(\cite{HAW86}, \cite{HAWx93})
fairly directly.
For the smallest threshold Hawkins applies (0.3~mag),
we extrapolate a $3\sigma$ upper limit of 0.012\% of the objects
varying that much.
This is more than two orders or magnitude smaller than his result.
Our variability fraction, even neglecting our lower threshold,
is approximately one tenth of his.
These results are also much smaller than
those of Trevese \etal
(\cite{TRE89}, \cite{TRE94})
who searched for
faint variable objects with $B < 22.6$ with a $\sigma>0.1 $~mag threshold
on photographic plates.
They found 64 variables out of the 694 objects with stellar PSF,
(seeing=$1.6\arcsec$)
from a total of nearly 1000 objects.
This implies a variable fraction of 9\%,
two orders of magnitude higher than our results,
but their restriction to objects with a stellar PSF
is expected to give them a sample somewhat enriched in AGNs.
Again, the threshold above which they will notice variability is higher than
ours, so they should see only a small fraction of the AGNs we find.
Our extrapolated $3\sigma$ upper limit is 0.15\% variable objects.
Both of those papers have investigated a sample of their objects
spectroscopically, and both found that they did a reasonably
good job of finding QSOs.

We see only two reasonable conclusions.
Either AGN variability changes dramatically near $B_j = 21$,
or prior searches have suffered from systematic errors that have
led them to spuriously claim non-variable QSOs to be variable.
There are few statistics on the distribution of optical variability
in AGNs.  Clearly, some vary significantly on 1-10 year time scales
%(Schramm \etal),
(\cite{SCHRa94}, \cite{SCHRb94}),
but most seem to be relatively constant.
For instance the Hamburg Quasar Monitoring Program finds that
most QSOs vary smoothly, typically by less than a tenth of a magnitude
per year
%(Borgeest\&Schramm).
(\cite{BOR94}).
Many of these quasars could not easily have been seen to be variable
by prior monitoring programs,
as the use of photographic plates forced variability-detection thresholds of
0.3~mag or more.
The only published claim that nearly all quasars vary dramatically is
based on the same Hawkins dataset
% (Hawkins\&V\'eron93).
(\cite{HAWV93}).

We believe that our CCD data is more reliable than previous
photographic searches.
Our practice of median-combining images from the same epoch eliminates defects
and reduces gain nonuniformities.
We also believe that our subtraction technique may cancel much of the 
systematic errors from which photographic plates suffer.
We note that the Sloan Digital Sky Survey will search for variables on a
very large area, at a slightly brighter magnitude range.
It should provide a useful check of our result.

\section{Variable Stars}

What fraction of our sample of variable objects could be stars?
Stars, particularly population I stars, are known to be variable.
In our galaxy's disk,
only intrinsically faint stars would be candidates,
because of our faint apparent magnitude range.
We integrate the luminosity function from
Kirkpatrick \etal
over a disk with a 350~pc scale height
(similar to that suggested by
the recent MACHO
%(Cook \etal 1994)
(\cite{COO94})
and OGLE
%(Paczynski \etal 1994)
(\cite{PAC94})
results),
and expect 20 disk stars in our magnitude range in the 2345+007 field.
The dominant contribution comes from M-dwarfs with $V\approx 12$,
which have a space density of $\approx 10^{-2}$~mag$^{-1}$~pc$^{-3}$;
such stars are relatively likely to be variable, as flare stars.
Our variable objects are divided roughly evenly between bluish objects
with $B_j - R \approx 0.6$ and red objects with $B_j - R \approx 2.3$;
the former group is rather unlikely to be a stellar population,
but it is not unreasonable to assume that the red half of our our variables
may be M-dwarfs.

Halo stars have much lower space density
($\approx 10^{-4}$~mag$^{-1}$~pc$^{-3}$),
but the halo is large, so that bright stars can contribute
from several kiloparsecs away.
The intrinsically brightest stars ($M_v \approx 9$) are the most common, 
as the sampled volume is so large;
it is the termination of the main sequence at $\approx 0.7$M$_\odot$ that
limits the sampled volume, rather than the size of the halo.
The total number of stars, using the \cite{DAH95}
%(C. C. Dahn \etal (in press))
luminosity function in our field is $\approx 250$.
Unlike disk stars, very few halo stars are variable, due to their
age and low metallicity
%(Boeshaar, 1995).
(\cite{BOE95}).
While we cannot accurately estimate the number of variable stars
in our field,
the possibility that a subset of our variable objects
may contain some stars makes our upper limits stronger,
and even less in agreement with previous photographic surveys.

\section{Predictions of microlensing}

Microlensing from dark matter has been tested in a variety of mass
ranges
%(See Dalcanton \etal 1994, Loeb 1994, and Press \& Gunn 1973).
(\cite{DAL94}, \cite{LOEx94}, \cite{PRE73}).
Our data can provide microlensing constraints,
but they are generally weaker than produced by the MACHO survey
%(Cook \etal).
(\cite{COO94}).
Microlensing of AGNs or quasars is difficult to disentangle because
the constraints contain equal amounts of information about
quasar structures as they do about the lens.
The mass range to which we are most sensitive is
$10^{-4} M_\odot$-$10^{-1} M_\odot$,
for AGNs which have a milliparsec hotspot.
We can show that this mass range contributes less than
$\Omega_r < 0.1$, assuming that 1\% of objects have milliparsec cores,
or alternatively, we can show that less than $0.1\%/\Omega_r$ of galaxies in
our magnitude range have milliparsec hot spots.

The importance of microlensing to the AGN phenomenon
has been suggested for some time
%(Vietri \& Ostriker 1983; Vietri 1985; Ostriker \& Vietri 1986;
%Schneider, 1986; Nottale, 1986),
(\cite{VIE83}, \cite{VIE85}, \cite{OST86}, \cite{SCH86}, \cite{NOT86})
though generally from a theoretical perspective.
Recently, Hawkins (1993) has claimed such an
interpretation for the variability of 18-22~mag galaxies he finds in
photographic photometry of patrol plates taken over a decade.
% Baganoff and Malkan (1995)
\cite{BAG95}
provides an opposing view;
see also
%Lacey (1994)
\cite{LAC94}
and \cite{SCH93}
for statistical discussions of microlensing and quasar variability.

\section{AGN structure}

The nearly-standard model of an AGN is an accretion disk surrounding a
black hole, with energetic particle beams along the rotation axis,
and a dusty torus in the plane of the disk (reviewed by
%Antonucci 1993).
\cite{ANT93}).
The observed variability will be a strong function of the viewing angle.
As yet, there is little information on what fraction of the
AGNs are oriented to give us an end-on view of the accretion disk,
and what fraction present us with the dust clouds
(with light scattering through the dust).
A crude approximation can be obtained by comparing the density of
BL Lacetae objects,
which are presumably AGNs viewed end-on
%(Perlman \etal 1995)
(\cite{PER95}).
to quasars or AGNs.
In a magnitude limited sample,
derived from the ESO catalog
%(V\'eron-Cetty and V\'eron 1993),
(\cite{VER93}),
we find that the BL-Lac to QSO or AGN ratio is near 1\%
(See
%Brown and March\~a
\cite{BRO93}
for a discussion of the uncertainties of BL Lacertae counts).

Our data can place some constraints on this model.
We can expect to see variability from a ``bare'' accretion disk;
the time scales for light travel time across the disk is short.
However, an obscured accretion disk is another matter.
Even with small obscuration (\eg 1~mag), the variability of
the source can disappear if the obscuring region is more than a few
parsecs across.
With a large obscuring region, photons will scatter,
and will take paths with perhaps years of extra time delay.
This will smooth out the variability by averaging the light
curve with many different time delays.
Similarly,
if we are not seeing any accretion disk light directly,
but instead we are seeing photons generated from axial beams,
we will not see any variability if the scattering out of the beam
occurs over a region larger than a few tens of parsecs.

If we make the reasonable assumptions that all accretion disks are
intrinsically variable
%(Mineshige and Shields, and Clarke and Shields),
(\cite{MIN90}, \cite{CLA89}),
and that obscured accretion disks
don't vary on our time scale,
then we will just see BL Lac-like objects as variable.
We would thus expect substantial variability in
0.01\%-0.05\% of objects - a small fraction of the total AGNs.
This number is comparable to what we observe, given the uncertanties
in the estimates,
and helps rationalize our observation that so few objects vary.

It should be possible to use the exponent, $\zeta$,
that describes the shape of the probability distribution of variability
(equation \ref{agnp_eq}),
to get information about the structure of the scattering regions around
the accretion disk in AGNs.
If one is willing to make assumptions about the
distribution of variability in accretion disks,
it may be possible to calculate how long photons must be delayed in the
scattering regions in order to produce the observed distribution
of variabilities.

\section{Conclusion and Summary}

In the first large, deep, multi-year CCD search for variable objects,
we have turned up far fewer candidates than found by previous
photographic searches.
We have confirmed this by median-combining
a set of photographic plates to form nearly defect-free images of
another area,
and have obtained a consistent, small, number of candidates. 

For $B_j \approx 22$ objects,
$0.74\% \pm 0.2 \%$ are AGNs that vary by 0.026~mag or more,
RMS, over the period of observation.
We find that the probability of a galaxy having a variability
$\delta M$ (near 0.026~mag) varies as ${(\delta M)}^{-3.3\pm 0.8}$.

\acknowledgements

We acknowledge the help of the expert staff at CTIO, where most of the
CCD data were acquired, and the assistance of Ata Saradjenini at KPNO.
Pat Boeshaar, Carol Christian, and Pat Waddell helped with the CFHT runs.
Gary Bernstein allowed us to use some data.
We gratefully acknowledge Rick Wenk for help in the reduction
and calibration of the Sextans data,
Marc Postman for scanning those plates,
Rogier Windhorst for the Sextans plate acquisition.
Ethan Vishniac and Michael Redmond for helpful comments.
Support for this work was provided by NASA through grant \# HF-01069.01-94A
from STScI which is operated by AURA, Inc. under NASA contract NAS5-26555.

\appendix

\section{Algorithms for Matching PSFs}\label{appAlg}

The {\it NLFS} program that fits together and subtracts images operates as
follows:

First, since the program is used in a highly automated routine,
there are a number of sanity checks on the data and the initial parameters,
so that the program will not produce silly results from silly data.

There is an outer loop that calls the fitting routine, progressively allowing
more parameters to be variable with each pass.
A luminosity scale and the sky levels are freed first,
followed by 
the coordinate transformation,
then a single parameter to specify the smoothing scale,
and finally the other smoothing parameters.

The fitting routine is a Marquardt algorithm for weighted least squares fitting.
Weights are set by the Poisson noise for each pixel.

However, before the fitting routine can be allowed to run, the data must be prepared.
The routine will not tolerate a variable number of data,
so bad pixels (marked by IEEE-754 Not-A-Numbers in the incoming data)
are  identified, and a mask is made that specifies that those and nearby pixels
will be ignored.
Similarly, pixels near the edge are masked off.
The mask is generated generously, so that invalid data will not be used,
even as the parameters
(\eg the coordinate transformation)
change during the fitting.

Finally, we identify blank sky, and mask off most of it, so that we really
only pay attention to the parts of the image near objects.
This is mostly a time- and memory- saving technique (we can thus ignore
90\% of the image), but also makes NLSFIT safer.
Safer, because the errors are not dominated by Poisson noise from the
blank sky (which contains no useful information), thus we are less
dependent on precise knowledge of correlations in the noise of
neighboring pixels.
Safer also, because the algorithm becomes less sensitive to approximations
we make in calculating the correct convolution kernels.

Inside the fitting routine, we calculate differences between the two
datasets, on a grid of output pixels (with some masked off holes).
The output coordinate system is chosen to be intermediate between
the two input systems.
Care must be taken to have a smooth and well-defined phase relationship
between the input and output coordinate systems,
otherwise shifting Moir\'e fringes will spoil the smooth
relationship between parameters and the error measure that the
fitting routine depends on.

The differences between the two datasets are calculated by
convolving each dataset by an appropriate kernel,
then interpolating each to the output coordinate system,
and subtracting.

The heart of the program, though, is the calculation of
the convolution kernels.
From some of the parameters, we construct a kernel, $\kappa$, that shows how
much we will smooth one image, {\it relative to how much we smooth the other}.
If the input images are $I_1$ and $I_2$,
and the output images are $O_1$ and $O_2$,
then $O_1 = I_1 \cdot q$ and $O_2 = I_2 \cdot q \cdot \kappa$,
where $q$ is also a kernel.
We find $q$ from the constraint that the noise in the final (sum or difference)
image must be white in order to fulfill the assumptions of the fitting routine.
If our output image is $O = O_1 \pm \alpha O_2$, then the noise in $O$ is
$\tilde{O} = \tilde{O_1} + \alpha^2 \tilde{O_2}$, where $\tilde{O}$ is the noise autocorrelation
function (or noise spectral power density) of $O$.
This may be expanded to yield
$\tilde{O} = \tilde{I_1} \cdot q^2 + \alpha^2 \tilde{I_2} q^2 \cdot \kappa^2$,
and solved in Fourier space for $q$:
$q^2 = \tilde{O} \cdot (\tilde{I_1} + \alpha^2 \tilde{I_2} \cdot \kappa^2)^{-1}$.
We can then require that $\tilde{O}$ be white and normalized to have unit variance:
$1 = \sum{\tilde{O}}$.
This has the nice feature that the total sum-squared errors on a noise image
(or one where the subtraction of two objects is perfect) is constant and unity;
anything above that is due to a misfit between the two images.

In general, these kernels must be truncated, so that convolution operations
can be done on real space.
Convolution in Fourier space makes it impossible to handle isolated bad pixels
(\eg saturated stars), and makes it difficult to handle images where the
sky background is not precisely flat.
Additionally, real space colvolutions are faster, if the kernel is sufficiently
(smaller than about 7x7).
Since the kernel size cannot be changed inside the least-squares fitting
routine, one of the functions of the outermost loop is to estimate a suitable
kernel size, based on the previous iteration.

The input noise autocorrelation functions ($\tilde{I_1}$ and $\tilde{I_2}$)
are functions of the fitting parameters through the pixel scale of the
images (more generally, the coordinate transformation matrix).
Imagine fitting a HST image to a typical image from a ground-based telescope.
The HST image has a pixel scale \eg four times finer than the other image.
In {\it NLSFIT}, the output image would have an intermediate pixel scale
(\eg twice as fine as the ground-based image), so each ground-based
pixel would be used twice.
When that image is interpolated to the scale of the output image,
before convolution and subtraction, adjacent pixels will then be highly
correlated.
On the HST image, we are seriously undersampling, and there would be no
correlation between pixels at all.

The actual parameters in the software unfortunately do not bear any simple
relationship to either the smoothing or sharpening kernels.
To keep the fitting routines operating smoothly,
we found it necessary to arrange a functional form where $\chi^2 $ is a
smooth function of all the fitting parameters,
with continuous first derivatives,
including the point where the images have identical PSFs.
To a first approximation, though,
the smoothing kernel is the sum of an elliptical Gaussian of variable
size, and an elliptical exponential PSF of the same size and orientation.
The exact functional form is not critical, as the kernel here is only supplying
difference between two PSFs (which are approximately the same size).
The relevant parameters are the size,
the relative amounts of Gaussian and
exponential (\ie long tailed or non-Gaussian shape),
and two terms to specify the elipticity and orientation.

\section{Matching 2345+007 and Sextans datasets}\label{appmatch}

We have only weak information for the Sextans
field on whether a given variable object is an AGN or a SN.
Since there are only two epochs, we cannot discriminate
on the shape of the light curve.
However
photometry (and all the analysis) is done on both a tight
aperture (the size of the PSF) and a larger one (3 times the size
of the PSF).
Objects are classified as to whether their variability is
more significant with the large or small aperture.

As AGNs are generally in the center of galaxies,
the small aperture should pick up all the signal and a minimal
amount of sky noise.
Thus, faint AGNs dominated by sky noise should be statistically more
significant in the small than the large aperture.
Supernovae will typically be slightly off-center in a galaxy, and thus will,
sometimes be more significant in the large aperture than the small one,
especially for $z < 0.2$.
Rather than attempting detailed simulations that depend upon unknowns such
as the spatial distribution  of SNe and black holes in galaxies
at high redshift, we finessed the  problem by including
the classification probabilities for the Sextans dataset
as model parameters.

The maximum likelihood procedure gives these two matching parameters properly
wide uncertainties, but properly propagates the small amount of information
from the spatial distribution into the rest of the parameters.
It also propagates information on AGN-to-SN ratios from
the 2345+007 data back into the Sextans dataset.
Pragmatically speaking, the above paragraph only has a noticeable effect on
the final SN to AGN ratios, a parameter without sufficient
signal-to-noise ratio to make any meaningful claims.

{}

\newpage

\begin{figure}
%\plotone{figs/fig1/2345.ps}
\caption{
A color image of the central region of the Q2345+007 field,
reconstructed from 26~hours of $B_j$ and R CCD exposures.
North is up and east is left;
the image is 200\arcsec across.\label{color}}
\end{figure}

\begin{figure}
% \plotone{figs/plate23/001.ps}
%\plotfiddle{figs/plate23/001.ps}{5.0in}{-90.}{90.}{-51.9}{-420}{150}
\caption{
The changes in the Q2345+007 field in $B_j$
between the 1987 and the 1990 epochs.
The point-spread functions of the two epochs have been matched
by {\it NLSFIT} before subtraction.
\label{diffimg}}
\end{figure}

\begin{figure}
%\plotone{figs/plate23/000.ps}
%\plotfiddle{figs/plate23/000.ps}{5.0in}{-90.}{90.}{-51.9}{-420}{150}
%\plotfiddle{figs/plate23/990J_out_kernel.ps}{0.0in}{0.}{20.}{20.}{-180}{0}
%\plotfiddle{figs/plate23/987J_out_kernel.ps}{0.0in}{0.}{20.}{20.}{30}{30}
\caption{
The sum of the $B_j$ band images of the
1987 and the 1990 epochs, processed through {\it NLSFIT}.
The imaged region is the same as above;
the grey scale is $asinh(signal/noise)$ (\ie approximately logarithmic).
Image addition was done by {\it NLSFIT};
it convolves the two images with kernels chosen to produce
white sky noise (\ie pixels are uncorrelated) in the
sum and difference images.
The two convolution kernels
are shown in the insets, with the left operating on the 1990
data and the right on the 1987 data before addition or subtraction.
The final PSF of the sum image is roughly intermediate between
the two input PSFs, but is not guaranteed to be Gaussian, and
can contain some oscillatory structure, as is seen here.
\label{sumimg}}
\end{figure}

\begin{figure}
%\plotone{figs/photerr/photerrG.ps}
\caption{
Photometry errors for 2345+007 (solid line) and Sextans (dashed line) fields
as a function of magnitude for the small aperture search
(aperture size equals PSF size).
Poisson noise (photon statistics) limit the noise for objects fainter
than $\approx 22$ mag,
and saturation of the photographic plates is significant for objects
brighter than $19$ mag.
The variability-detection thresholds for individual objects are shown as dots.
The threshold changes from object to object
depending on the number of overlapping frames in which a particular
object was imaged,
so the variability-detection thresholds of uniformly imaged areas
areas tend to form smooth curves on this plot.
Both Sextans and 2345+007 variability-detection thresholds are
plotted for both small aperture and large aperture searches.
Objects that were actually detected as variable are shown as hexagons,
with their measured variability (true variability plus noise)
plotted as the y-axis.
Clearly, there are few objects with strong variation;
many are marginal detections.
\label{photerrG}}
\end{figure}

\begin{figure}
%\plotone{figs/photerr/photerrB.ps}
\caption{
Photometry errors for 2345+007 (solid line) and Sextans (dashed line) fields
as a function of magnitude for the large aperture search.
Poisson noise (photon statistics) limit the noise for objects fainter
that $\approx 21$ mag,
and saturation of the photographic plates is significant for objects
brighter than $20$ mag.
The variability-detection threshold (individual dots) are several times higher,
depending on the number of overlapping frames in which a particular
object was imaged.
\label{photerrB}}
\end{figure}

\begin{figure}
% \plotfiddle{figs/histogram/histogram.ps}{5.0in}{0.}{50.}{50.}{0}{0}
%\plotfiddle{figs/histogram/histogram.ps}{5.0in}{90.}{70.}{70.}{300}{0}
\caption{
Histograms of the distribution of chi-squared
statistics for the Sextans objects (upper) and blank sky spots
(lower).
The number of objects between the curves
is the number of variable objects.
Some objects (\eg near $-log_{10} ( P_{false} ) = 4$ )
can be used statistically,
but we cannot be certain which of those objects are variable.
The 2345+007 histogram is
similar, but has fewer surveyed objects and correspondingly fewer variables.
\label{histogram}}
\end{figure}

\begin{figure}
% \plotfiddle{figs/classprob/classprob.ps}{5.0in}{0.}{50.}{50.}{0}{0}
%\plotfiddle{figs/classprob/classprob.ps}{5.0in}{90.}{70.}{70.}{300}{0}
\caption{
The classification (including four misclassifications)
probabilities for the 2345+007 dataset.
The right axis is the probability of detecting a type-1a supernova;
it is small because a supernova can flare and fade in the $\approx$1~year
interval between images.
The top curve for the left axis is the probability
of detecting a SN and classifying it as an SN, as a function of the magnitude
of the SN.
The left axis is the probability of detecting an AGN to be variable,
and the top curve is the probability of correctly classifying it as an AGN.
AGN probabilities are plotted against the average magnitude of the stellar
nuclear component, with an assumed variability of $\delta L / L = 1$.
Arrows on the figure show which axis to use.
These probabilities are shown for faint objects ($B_j<21$) where the
signal to noise ratio is limited by photon statistics from the sky
background.
\label{classprob}}
\end{figure}

\begin{figure}
%\plotone{figs/colormag/colormag.ps}
\caption{
The 2345+007 variable objects plotted on a color-magnitude
diagram.
Variable objects are shown as a ``+'' ($B_j$) or ``x'' ($R$),
to allow comparison.
Circled points are variable in the other color with lower
(99\%) confidence.
It can be seen that there is substantial commonality,
suggesting that these objects are indeed variable.
A substantial fraction of the very bluest objects are seen to be variable,
and are likely QSOs.
\label{colormag}}
\end{figure}

\begin{figure}
%\plotone{figs/sizemag/sizemag.ps}
\caption{
The 2345+007 variable objects plotted on a size-magnitude
diagram.
Size is plotted as the square of the FWHM (full-width at half max) of
a Gaussian fit to the object,
with seeing subtracted.
Variable objects are shown as a ``+'' ($B_j$), or ``x'' ($R$),
to allow comparison.
Circled points are variable in the other color with lower
(99\%) confidence.
Many of the variable objects are unresolved.
The sequence of unresolved objects is particularly apparent for
bright objects (\ie brighter than $20^{th}$ mag,
and size-squared$ = 0 \pm 0.1 \arcsec^2$.
\label{sizemag}}
\end{figure}

\begin{figure}
%\plotone{figs/closeups/allJ.ps}
\caption{
$B_j$ band images of variable and possibly variable objects in 2345+007.
Each image is $18 \arcsec$ across, centered on the object,
and is a composite of as many epochs as are available at that position.
The seeing and noise thus vary somewhat from image to image.
Intensity
is plotted as $asinh(Brightness/Noise)$, which produces a nice logarithmic
scale for bright regions, yet behaves smoothly near zero (sky)
brightness.
% These images are constructed using all available epochs, and without any bad
% pixels excluded.
% G7676 shows some bad pixels from a defective CCD column in the 1994 epoch;
% unlike in these images,
% bad pixels were excluded, epoch by epoch, from the analysis procedure.
The list of objects used is a composite of all detections in either the
$B_j$ or $R$ searches.
\label{closeupJ}}
\end{figure}

\begin{figure}
%\plotone{figs/closeups/allR.ps}
\caption{
$R$ band images of variable and possibly variable objects in 2345+007. 
Intensity
is plotted as $asinh(Brightness/Noise)$.
Each image is $18 \arcsec$ across, centered on the object.
Images are paired, $B_j$ and $R$ bands, with the preceding figure.
\label{closeupR}}
\end{figure}

\begin{figure}
%\plotone{figs/closeups/allS.ps}
\caption{
$B_j$ band images of variable objects in Sextans.
Intensity
is plotted as $asinh(Brightness/Noise)$.
Each image is $18 \arcsec$ across, centered on the object.
% Note the rarity of plate defects;
% only G3337 is under a ghost image,
% and G645, G653, G686, and G710 are under a glint
% from an off-plate star on one epoch.
\label{Scloseup}}
\end{figure}

\newpage

\begin{deluxetable}{lcccccl}
\tablecolumns{7}
\tablewidth{0pc}
\tablecaption{Journal of Observations\label{journal}}
\tablehead{
\colhead{Date} &
\colhead{Observatory} &
\colhead{CCD} &
\colhead{Scale} &
\colhead{Filter} &
\colhead{Exp.} &
\colhead{FWHM} \nl
\colhead{(dd/mm/yy)} &  &  & \colhead{\arcsec / pix}
&  & \colhead{(sec)} & \colhead{\arcsec } \nl
}
\startdata
24/10/84 - 25/10/84 & CTIO & RCA $~320 \times ~508$ & 0.59~ & B$_j$ & $~7 \times 900$s & 1.6 - 1.9 \nl
23/10/84            & CTIO & RCA $~320 \times ~508$ & 0.59~ & B$_j$ & $~1 \times 500$s & 1.5 \nl
23/10/84            & CTIO & RCA $~320 \times ~508$ & 0.59~ & R$_{~}$ & $11 \times 500$s & 1.2 - 1.3 \nl
% 23/10/84 - 24/10/84 & CTIO & RCA $~320 \times ~508$ & 0.59~ & I     & $14 \times 500$s & 1.1 - 1.4 \nl
\tablevspace{6pt}
09/11/85 - 10/11/85 & CTIO & RCA $~320 \times ~508$ & 0.59~ & B     & $~8 \times 500$s & 1.3 - 1.6 \nl
\tablevspace{6pt}
17/09/87 - 19/09/87 & CFHT & RCA ~$340 \times ~528$ & 0.41~ & B$_j$ & $31 \times 500$s & 0.8 - 1.1 \nl
17/09/87 - 19/09/87 & CFHT & RCA $~340 \times ~528$ & 0.41~ & R     & $11 \times 500$s & 0.9 - 1.1 \nl
\tablevspace{6pt}
24/10/89 - 25/10/89 & CFHT & RCA $~340 \times ~512$ & 0.41~ & B$_j$ & $~9 \times 500$s & 0.8 - 1.0 \nl
24/10/89 - 25/10/89 & CFHT & RCA $~340 \times ~512$ & 0.41~ & R     & $~9 \times 500$s & 0.7 - 1.0 \nl
\tablevspace{6pt}
18/08/90            & KPNO & Tek $1024 \times 1024$ & 0.47~ & B$_j$ & $~3 \times 300$s & 1.2 \nl
17/08/90            & KPNO & Tek $1024 \times 1024$ & 0.47~ & R     & $~3 \times 300$s & 1.0 - 1.1 \nl
\tablevspace{6pt}
23/09/90 - 24/09/90 & CTIO & TI~ $~800 \times ~800$\tablenotemark{*} & 0.58~ & B$_j$ & $20 \times 300$s & 1.2 - 1.5 \nl
23/09/90 - 24/09/90 & CTIO & TI~ $~800 \times ~800$\tablenotemark{*} & 0.58~ & R     & $25 \times 300$s & 1.1 - 1.3 \nl
\tablevspace{6pt}
08/09/91            & KPNO & Tek $1024 \times 1024$ & 0.47~ & B$_j$ & $~3 \times 300$s & 1.2 - 1.3 \nl
08/09/91            & KPNO & Tek $1024 \times 1024$ & 0.47~ & R     & $~3 \times 300$s & 1.0 - 1.1 \nl
\tablevspace{6pt}
08/12/91            & KPNO & Tek $2048 \times 2048$ & 0.47~ & R     & $~3 \times 720$s & 1.3 - 1.4 \nl
\tablevspace{6pt}
26/10/92 - 27/10/92 & CTIO & Tek $1024 \times 1024$ & 0.48~ & B$_j$ & $12 \times 500$s & 1.3 - 1.7 \nl
26/10/92 - 27/10/92 & CTIO & Tek $1024 \times 1024$ & 0.48~ & R     & $13 \times 400$s & 1.2 - 1.6 \nl
\tablevspace{6pt}
23/06/93            & CTIO & Tek $2048 \times 2048$ & 0.48~ & B$_j$ & $~6 \times 500$s & 1.2 - 1.5 \nl
\tablevspace{6pt}
11/01/94            & CFHT & Loral $2048 \times 2048$ & 0.21 & R & $~3 \times 600$ & 0.7 \nl
\tablevspace{6pt}
30/08/94            & KPNO & Tek $2048 \times 2048$ & 0.47~ & B$_j$ & $~6 \times 500$s & 1.3 - 1.4 \nl
30/08/94            & KPNO & Tek $2048 \times 2048$ & 0.47~ & R     & $~6 \times 500$s & 1.3 - 1.4 \nl
\tablevspace{6pt}
24/09/94            & KPNO & Tek $2048 \times 2048$ & 0.47~ & B$_j$ & $~6 \times 500$s & 1.3 - 1.4 \nl
24/09/94            & KPNO & Tek $2048 \times 2048$ & 0.47~ & R     & $~6 \times 500$s & 1.3 - 1.4 \nl
\enddata
\tablenotetext{*}{Rebinned to $400 \times 400$}
\end{deluxetable}

\begin{deluxetable}{llcccccc}
\tablecolumns{8}
\tablewidth{0pc}
\tablecaption{Variable Objects in 2345+007\label{tabvar}}
\tablehead{
\colhead{Object} & \colhead{RA [2000]} & \colhead{DEC} & \colhead{mag} &
\colhead{$B_j-R$} & \colhead{size} & \colhead{No. images} &
\colhead{$\chi^2/N$} \nl
\colhead{ID} & \colhead{(hms)} & \colhead{(dms)} & \colhead{$(B_j)$} &
\colhead{(mag)} & \colhead{($arcsec^2$)} & \colhead{($B_j$,$R$)} &
\colhead{($B_j$,$R$)} \nl
}
\startdata
G1379 & $23^{h}~47^{m}~41.0^{s}$ & $0\deg~59\arcmin~3\arcsec$ & 23.6 & 0.8 & -0.4 & 2,~4 & 0.1,~16.5\nl
G2735 & $23^{h}~47^{m}~42.0^{s}$ & $1\deg~2\arcmin~58\arcsec$ & 22.2 & 1.6 & 0.1 & 2,~3 & 6.8,~17.3\nl
G386 & $23^{h}~47^{m}~38.7^{s}$ & $0\deg~54\arcmin~40\arcsec$ & 21.9 & 1.1 & 0.2 & 2,~3 & 0.0,~16.8\nl
G442 & $23^{h}~47^{m}~40.1^{s}$ & $0\deg~54\arcmin~28\arcsec$ & 22.6 & 1.8 & 0.2 & 2,~4 & 0.0,~17.2\nl
G5158 & $23^{h}~48^{m}~11.5^{s}$ & $0\deg~57\arcmin~0\arcsec$ & 22.0 & 0.4 & -0.3 & 6,~7 & 45.4,~9.1\nl
G5740 & $23^{h}~48^{m}~12.7^{s}$ & $0\deg~57\arcmin~49\arcsec$ & 23.7 & 0.4 & -0.1 & 9,~10 & 198.5,~33.1\nl
G5740 & $23^{h}~48^{m}~12.7^{s}$ & $0\deg~57\arcmin~49\arcsec$ & 23.7 & 0.4 & -0.1 & 9,~10 & 198.5,~33.1\nl
G5876 & $23^{h}~48^{m}~6.5^{s}$ & $1\deg~0\arcmin~42\arcsec$ & 22.0 & 0.5 & 0.5 & 2,~6 & 13.7,~4.4\nl
G6659 & $23^{h}~48^{m}~16.5^{s}$ & $0\deg~58\arcmin~16\arcsec$ & 23.7 & 2.1 & -0.3 & 10,~11 & 3.8,~11.4\nl
G6697 & $23^{h}~48^{m}~19.1^{s}$ & $0\deg~57\arcmin~17\arcsec$ & 20.6 & -0.0 & -0.3 & 10,~11 & 15.4,~12.8\nl
G6789 & $23^{h}~48^{m}~19.5^{s}$ & $0\deg~57\arcmin~20\arcsec$ & 19.3 & -0.1 & -0.4 & 10,~11 & 43.9,~23.1\nl
G6789 & $23^{h}~48^{m}~19.5^{s}$ & $0\deg~57\arcmin~20\arcsec$ & 19.3 & -0.1 & -0.4 & 10,~11 & 43.9,~23.1\nl
G7676 & $23^{h}~48^{m}~11.8^{s}$ & $1\deg~3\arcmin~0\arcsec$ & 23.3 & 0.1 & -0.2 & 2,~3 & 1.0,~18.8\nl
\enddata
\tablecomments{ This table shows objects detected with the small aperture search, in either $B_j$ or $R$ bandpasses. The color is the central color, resulting from the small aperture search. The size is the FWHM of the central peak, with seeing quadratically subtracted out, as derived from moments in the photometric aperture (negative numbers reflect errors in FWHM measurements). The number of images column notes how many of the observing runs imaged each variable object in each color. G6697 and G6789 are the quasar A and B images, respectively. }
\end{deluxetable}

\end{document}